\documentclass{roajarticle}
\usepackage{graphicx}
\usepackage{mynatbib}
\usepackage{upgreek}
\usepackage{upgreek2}
\usepackage{amsmath}
\usepackage{float}
\usepackage[section]{placeins}
\newcommand{\volume}{\textbf{1}}
\newcommand{\numarul}{1}
\newcommand{\volyear}{2022}
\graphicspath{ {folder} }

\title{Dark matter and radiation production during warm inflation in a curved Universe-an
	irreversible thermodynamic approach}
\newcommand{\shorttitle}{Dark matter and radiation in warm inflation}

\author[1]{Teodora  \MakeTextUppercase{Matei}}
\author[2]{Tiberiu  \MakeTextUppercase{Harko}}
\author[3] {Gabriela \MakeTextUppercase{Mocanu}}

\affil[1]{Department of Physics, Babes-Bolyai University,
	1 Kogalniceanu Street, Cluj-Napoca, 400084, Romania,
	Email: teodora.maria.matei@stud.ubbcluj.ro}

\affil[2]{Department of Theoretical Physics, National Institute of Physics
	and Nuclear Engineering (IFIN-HH), Bucharest, 077125 Romania,
	Email: tiberiu.harko@aira.astro.ro}

\affil[3]{Romanian Academy, Astronomical Observatory, 19 Ciresilor Street,
	Cluj-Napoca, 400487, Romania,
	Email: gabriela.mocanu@aira.astro.ro}

\keywords{Cosmology; Warm inflation; Irreversible thermodynamics; Curvature effects}

\begin{document}
	\maketitle
	
	\begin{abstract}
		We investigate the creation of dark matter particles as a result of the decay of
		the scalar field in the framework of warm inflationary models, by using the
		irreversible thermodynamics of open systems with matter
		creation/annihilation. We consider the scalar fields, radiation and dark
		matter as an interacting three component cosmological fluid in a
		homogeneous and isotropic Friedmann-Lemaitre-Robertson-Walker (FLRW) Universe, in the presence of the curvature terms. The thermodynamics of
		open systems as applied together with the gravitational field equations to
		the three component cosmological fluid leads to a generalization of the
		elementary scalar field-radiation interaction model, which is the
		theoretical basis of warm inflationary models. Moreover, the decay
		(creation) pressures describing matter production are explicitly considered
		as parts of the cosmological fluid energy-momentum tensor. A specific
		theoretical model, describing coherently oscillating scalar waves, is considered. In particular, we investigate the role of the curvature terms in the dynamical evolution of the early Universe, by considering numerical solutions of the gravitational field equations. Our results indicate that despite the fact that the Universe becomes flat at the end of the inflationary era, the curvature terms, if present, may still play an important role in the very first stages of the evolution of the Universe.
		
	\end{abstract}
	
	
	\section{Introduction}
	
	The question of the homogeneity settled over far apart regions in space, respectively the queries concerning the horizon and the flatness problems of the Universe, are beautifully answered by the theory of inflation, introduced in \citet{Guth1981}. Alan Guth’s  “old inflation” requires the existence of a scalar field, whose energy-momentum tensor mimics that of an ideal fluid. For a detailed discussion of the properties of scalar fields see \citet{Mukhanov2005} and \citet{Nojiri2017}, respectively. Later on, a “new inflation” scenario was proposed, in which the self-interacting potential $V(\upphi)$ of the scalar field was set to be nearly flat at its minimum, where it undergoes oscillatory fluctuations \citep{Linde1982, Albrecht1982}. Due to multiple complications raised by this model, the chaotic inflationary scenario was developed in \citet{Linde1983} and \citet{Linde1994}, respectively. In the chaotic inflation model one considers a region in space where at the initial time $t=t_{0}$ the scalar field is very large, and approximately homogeneous. The energy-momentum tensor of the scalar field is dominated by the large potential, thus leading to an equation of state $p_\upphi\approx -\uprho_\upphi$, and to an inflationary expansion.
	
	The above mentioned inflationary models are \textit{“cold inflation”} models, in which the amplitude of the oscillating scalar field decreases, due to the expansion of the Universe, and because of an energy loss. As the Universe enters a cold, low-energy phase, a reheating phase, describing the production of Standard Model particles after the phase of accelerated expansion, is necessary, with the elementary particle production reaching thermal equilibrium at a certain reheating temperature. A slow reheating was effectively studied in chaotic inflation, \citep{Albrechts1982, Kofman1994, Antusch2015}. Hence, the slow-roll inflation describes the early exponential expansion of the Universe and the reheating phase as two different processes.
	
	The remnants of inflation can be observed in the temperature anisotropies of the Cosmic Microwave Background Radiation, by measuring the temperature fluctuations due to the  variations of the matter density throughout the universe. The theoretical models in which the density perturbations have their origin in the quantum fluctuations of the scalar field, are known to run into some difficulties, as pointed out, for example, in \citet{Berera1995}.  Henceforth, a novel theory of inflation emerged, starting with the works by \citet{BereraFang}, \citet{Berera1995} and \citet{BereraMoss}, respectively, namely, the \textit{“warm inflation”} scenario, in which the scalar field interacts dynamically with other fields through thermal dissipation, thus removing the quantum problem. Therefore, in the warm inflation proposal,  the thermal fluctuations dominate over the quantum disturbances \citep{BereraFang}.
	
	It was pointed out that warm inflation agrees neatly with COBE’s measurements of temperature anisotropies in the CMB radiation \citep{Berera1996}, as opposed to the reheating model, which is not as consistent with observational data. In the warm inflationary scenario the transition from the scalar field-dominated Universe to the radiation era is accomplished by a dissipation coefficient $\Upgamma$, inserted in the scalar field’s equation of motion, which accounts for a continuous energy transfer through its decay. The scalar field and the radiation form together a cosmological fluid, for which the entropy perturbations have been studied in \citet{Oliveira2001}. Various aspects of the warm inflationary Universe have been considered throughout the literature in \citet{Gleisler1994}, \citet{Hall2004}, \citet{Liddle1994}, \citet{Stewart2002}, and \citet{Bastero2011}, respectively.
	
	In the inflationary models the scalar field decays into matter, and thus particles are created. For a long time this process was investigated almost exclusively quantum mechanically. On the other hand, a thermodynamic approach to the adiabatic production of particles was introduced in \citet{Prigogine1988}, and further developed in \citet{Calvao1992} and \citet{Zimdahl1996}, respectively,  where a constant entropy per particle was imposed during matter production. The cosmological fluid in the presence of particle creation was described thermodynamically in \citet{Prigogine1988}, where the role played by the entropy in this process was pointed out. The decaying/creation process is assumed to be irreversible, allowing only for an energy flow from the scalar field to the newly created matter constituents. Henceforth, the particle creation in the framework of the irreversible thermodynamics of open systems has been applied for the study of various cosmological processes in \citet{Qiang2007}, \citet{Modak2012}, \citet{Harko2013}, \citet{Chakraborty2014}, \citet{ChakrabortySaha}, \citet{Lima2014}, \citet{Harko2014}, \citet{Nunes2016}, \citet{Lima2016}, \citet{Pigozzo2016}, \citet{Su2017}, \citet{Harko2020}, \citet{Harko2021}, and \citet{Harko2022}, respectively. Thus, in \citet{Harko2013}, the interaction between dark matter and dark energy was modeled by using an approach based on the thermodynamics of irreversible processes, while in \citet{Su2017}, the reheating theory was investigated by considering a three-component cosmological fluid.
	
	The interaction between scalar fields and radiation in the framework of warm inflationary models was considered, for a homogeneous, spatially flat and isotropic FLRW Universe, in \citet{Harko2020}. By assuming that the scalar fields and radiation form an interacting two component cosmological fluid, one can apply to this system 
	the physical formalism of the irreversible thermodynamics of open systems in the presence of matter creation/annihilation. The theoretical predictions of the warm inflationary scenario with irreversible particle production were also compared with the Planck 2018 observational data, and thus constraints on the free parameters of the model were obtained.
	
	It is the goal of the present paper to extend the warm inflationary scenario in three directions. First, since particle creation/decay processes represent dissipative irreversible processes, their proper formulation must be done in the framework of the thermodynamics of open system, as pointed out in \citet{Harko2020}. Secondly, we will consider that not only radiation, but also a dark matter component, interacting with the scalar field and the radiation, is created during the initial stages of the expansion. Thirdly, we will assume that the Universe was not flat at its birth, but it may have had an (arbitrary) curvature, and we will investigate the role and evolution of the curvature during the particle creation processes. The feedback of the created matter on the cosmological expansion will also be considered.
	
	The present paper is organized as follows. The basic formalism of the irreversible thermodynamics of open systems, together with the principles of warm inflation, are reviewed in Section \ref{sect2}. The full set of equations describing warm inflationary models with irreversible radiation and dark matter production due to the decay of the scalar field are written down in Section \ref{sect3}. As an application of the considered warm inflationary scenario in Section \ref{sect4} we consider the case of the coherently oscillating scalar field, which is analyzed by numerically integrating the system of the gravitational field equations.  We discuss and conclude our results in Section \ref{sect5}.
	
	\section{Thermodynamics and warm inflation} \label{sect2}
	
	In the present Section we will briefly review the fundamentals of the thermodynamics of open systems in the presence of particle creation, and we will also present the standard formulation of the warm inflationary cosmological scenario.
	
	\subsection{Brief review of irreversible thermodynamics with matter creation}
	
	The basic equations describing the thermodynamic of irreversible processes
	in the presence of matter creation have been developed in \citet{Prigogine1988}. In particular, the time variation of the matter energy density in systems with particle production is given by
	\begin{equation}
		\dot{\uprho} = \frac{h}{n}\dot{n},  \label{rho}
	\end{equation}
	where the overdot designates the derivative with respect to time, $h$ is the enthalpy per unit volume, defined as $h=\uprho+p$, where $\uprho$ and $p$ denote the energy density and the thermodynamic pressure, respectively. Another important concept in the theory of irreversible processes with matter creation is the creation pressure, defined as  \citep{Prigogine1988},
	\begin{equation}\label{pc}
		p_{c}=-\frac{h}{n}\frac{d\left( nV\right) }{dV}=-\frac{h}{n}\frac{V}{\dot{V}}%
		\left( \dot{n}+\frac{\dot{V}}{V}n\right) ,
	\end{equation}%
	where $V$ is the comoving volume, and $n$ denotes the particle number density, defined as $n=N/V$. The introduction of the creation pressure
	in the thermodynamical formalism represents an effective way to describe
	particle production, allowing the reformulation of the second law of
	thermodynamics for adiabatic systems in the presence of matter creation as \citep{Prigogine1988},
	\begin{equation}
		d\left( \uprho V\right) +\left( p+p_{c}\right) dV=0.
	\end{equation}
	
	In the following we will assume that the Universe can be described by a
	homogenous and isotropic geometry, given by the
	Friedmann-Lemaitre-Robertson-Walker (FLRW) metric, which has the general
	form
	\begin{equation}
		ds^{2}=c^{2}dt^{2}-a^{2}(t)\left[ \frac{dr^{2}}{1-kr^{2}}+r^{2}\left(
		d\uptheta ^{2}+\sin ^{2}\uptheta d\upphi ^{2}\right) \right] ,  \label{FLRW}
	\end{equation}%
	where $k=0,\pm 1$ represents the curvature of the space.  Dimensionally, $k$ has units of
	length$^{-2}$, $r$ has units of length, and $a(t)$ is dimensionless. An
	important cosmological quantity is the Hubble function, defined as $H(t)=%
	\dot{a}(t)/a(t)$. For the FLRW metric, the comoving volume is given by $%
	V(t)=a^{3}(t)$, giving $\dot{V}(t)=3a^{2}(t)\dot{a}(t)$. Therefore, for
	the creation pressure we obtain
	\begin{equation}
		p_{c}=-\frac{\uprho +p}{n}\frac{a}{3\dot{a}}\left( \dot{n}+\frac{3\dot{a}}{a}%
		n\right) =-\frac{\uprho +p}{3nH}\left( \dot{n}+3Hn\right) .
	\end{equation}%
	
	In order to close the thermodynamic equations we need a supplementary
	relation for the variation of the particle number, which in a cosmological
	framework can generally be taken as
	\begin{equation} \label{n}
		\dot{n}+3Hn=\Uppsi ,
	\end{equation}%
	where $\Uppsi $ is a source term. When $\Uppsi >0$, particle creation occurs,
	while $\Uppsi <0$ corresponds to matter/particle decay. Hence, the creation
	pressure is obtained as
	
	\begin{equation} \label{eq7}
		p_{c}=-\frac{\uprho +p}{3nH}\Uppsi .
	\end{equation}
	
	\subsection{Standard formulation of warm inflation}
	
	The warm inflationary scenario is an interesting and important theoretical alternative to the cold inflation and reheating theories \citep{BereraFang, Berera1995}. In the warm inflation scenario, the Universe experiences an accelerated very early expansionary stage, triggered by the presence of a scalar field. But, as opposed to the cold inflation model, besides the scalar field, a matter component of the cosmological fluid, usually assumed to be radiation, is also present, being generated by
	the decay of the scalar field. During the cosmological evolution, these two components interact dynamically. In its standard formulation, the cosmological evolution in warm inflation is described by the Friedmann equations
	\begin{equation}
		3H^{2} = \frac{1}{M_{P}^{2}}\left( \uprho _{\upphi }+\uprho _{rad}\right) ,
	\end{equation}%
	and
	\begin{equation}
		2\dot{H} = -\frac{1}{M_{P}^{2}}\left( \dot{\upphi}^{2}+\frac{4}{3}\uprho
		_{rad}\right) ,
	\end{equation}%
	respectively, where by $M_{P}=\sqrt{\hbar c/G}$ we have denoted the Planck
	mass, $\upphi $ is the scalar field, and $\uprho _{\upphi }$ and $\uprho _{rad}$
	represents the energy densities of the scalar field and radiation,
	respectively. The energy density $\uprho _{\upphi }$ and the pressure $p_{\upphi }$ of the scalar field are
	given by
	\begin{equation}
		\uprho _{\upphi }=\frac{\dot{\upphi}^{2}}{2}+V\left( \upphi \right) ,
	\end{equation}%
	and
	\begin{equation}
		p_{\upphi }=\frac{\dot{\upphi}^{2}}{2}-V\left( \upphi \right) ,
	\end{equation}%
	respectively, where $V\left( \upphi \right) $ is the self-interaction
	potential of the field.
	
	Due to the decay of the scalar field, which is essentially a dissipative
	process, energy is transferred from the field to the radiation fluid. In the standard
	warm inflationary scenario this process is described by the following energy
	balance equations \citep{BereraFang, Berera1995},
	\begin{equation}
		\dot{\uprho}_{\upphi }+3H\left( \uprho _{\upphi }+p_{\upphi }\right) =-\Upgamma \dot{\upphi%
		}^{2},  \label{eq1}
	\end{equation}
	\begin{equation}
		\dot{\uprho}_{rad}+3H\left( \uprho _{rad}+p_{rad}\right) =\Upgamma \dot{\upphi}^{2},
	\end{equation}%
	where $\Upgamma >0$ is the dissipation coefficient. By taking into account the
	explicit expressions of the scalar field energy density and pressure, Eq.~(%
	\ref{eq1}) takes the form of a Klein-Gordon type equation,
	\begin{equation}
		\ddot{\upphi}+3H\left( 1+Q\right) \dot{\upphi}+V^{\prime }\left( \upphi \right) =0,
	\end{equation}%
	where $Q=\Upgamma /3H$. In the following we will consider the natural system of units with $\hbar =G=1$, and $M_P^2=1$, respectively.
	
	\section{Particle creation in the warm inflationary scenario in a curved Universe} \label{sect3}
	
	In the following we consider a curved Universe, described by the general
	FLRW metric (\ref{FLRW}), and containing three basic components: a decaying
	scalar field, which generates both radiation and a (dark) matter component. We
	denote the particle number densities of these components as $n_{\upphi }$, $%
	n_{rad}$, and $n_{DM}$, respectively. The radiation fluid satisfies the
	equation of state $p_{rad}=\uprho _{rad}/3$, with $\uprho _{rad}=8\uppi
	^{5}T^{4}/15$, where $T$ is the temperature of the fluid. The photon number
	density is given as a function of the temperature by $n_{rad}=16\uppi
	\upzeta \left( 3\right) T^{3}$, where $\upzeta (n)$ is the Riemann zeta
	function. We also assume that dark matter is produced with a negligible
	pressure, $p_{DM}=0$, and therefore $\uprho _{DM}=m_{DM}n_{DM}$, where $m_{DM}$
	is the mass of the dark matter particle.
	
	In order to apply the formalism of the thermodynamics of open systems, we
	need to formulate first the particle balance equations, which originate from Eq.
	(\ref{n}). Moreover, in the following we assume that the source term $\Uppsi$ is proportional to the energy
	density $\uprho _{\upphi }$ of the scalar field, the proportionality coefficient being given by the dissipation coefficient $\Upgamma$, also giving
	the particle creation/decay rates. Hence, the balance equations of the particle number densities take the following form,%
	\begin{eqnarray}
		\dot{n}_{\upphi }+3Hn_{\upphi } &=&-\Upgamma \frac{\uprho _{\upphi }}{m_{\upphi }}, \\
		\dot{n}_{rad}+3Hn_{rad} &=&\frac{\Upgamma }{2}\frac{\uprho _{\upphi }}{m_{\upphi }},
		\\
		\dot{n}_{DM}+3Hn_{DM} &=&\frac{\Upgamma }{2}\frac{\uprho _{\upphi }}{m_{\upphi }},
	\end{eqnarray}%
	where $m_{\upphi }$ is the mass of the scalar field particle, and the
	dissipation coefficients have been chosen in such a way that the total
	particle number $n=n_{\upphi }+n_{rad}+n_{DM}$ is conserved. Once the particle
	balance equations are known, from Eq. (\ref{rho}) one obtains the energy
	densities of the newly created particles as
	\begin{equation}
		\dot{\uprho}_{\upphi }=\left( \uprho _{\upphi }+p_{\upphi }\right) \frac{\dot{n}_{\upphi
		}}{n_{\upphi }}=\dot{\upphi}^{2}\left( -3H-\Upgamma \frac{\uprho _{\upphi }}{m_{\upphi
			}n_{\upphi }}\right) ,  \label{KG}
	\end{equation}
	\begin{equation}
		\dot{\uprho}_{rad}=\left( \uprho _{rad }+p_{rad }\right) \frac{\dot{n}_{rad
		}}{n_{rad }}=\frac{4}{3}\uprho _{rad}\left( -3H+\frac{\Upgamma }{2}\frac{%
			\uprho _{\upphi }}{m_{\upphi }n_{rad}}\right) ,
	\end{equation}
	\begin{equation}
		\dot{\uprho}_{DM}=\left( \uprho _{DM }+p_{DM }\right) \frac{\dot{n}_{DM
		}}{n_{DM }}=\uprho _{DM}\left( -3H+\frac{\Upgamma }{2}\frac{\uprho _{\upphi }}{%
			m_{\upphi }n_{DM}}\right) .
	\end{equation}
	
	For the creation pressures, given by Eq. (\ref{eq7}), we obtain
	\begin{equation}
		p_{c}^{(\upphi )}=\Upgamma \frac{\dot{\upphi}^{2}}{3H}\frac{\uprho _{\upphi }}{%
			m_{\upphi }n_{\upphi }},
	\end{equation}
	\begin{equation}
		p_{c}^{(rad)}=-\frac{2\Upgamma }{3}\frac{\uprho _{rad}}{3n_{rad}H}\frac{\uprho
			_{\upphi }}{m_{\upphi }},
	\end{equation}
	\begin{equation}
		p_{c}^{(DM)}=-\frac{\Upgamma }{2}\frac{\uprho _{DM}}{3n_{DM}H}\frac{\uprho _{\upphi }%
		}{m_{\upphi }}.
	\end{equation}
	
	The basic equations describing the dynamical evolution of the scale factor
	in the warm inflationary curved Friedmann-Lemaitre-Robertson-Walker Universe
	filled with a decaying scalar field, and radiation and dark matter creation, respectively,
	are given by
	\begin{eqnarray}
		3\frac{\dot{a}^{2}}{a^{2}}+3\frac{k}{a^{2}} &=&\frac{1}{M_{P}^{2}}\left(
		\uprho _{\upphi }+\uprho _{rad}+\uprho _{DM}\right) , \\
		2\frac{\ddot{a}}{a}+\frac{\dot{a}^{2}}{a^{2}}+\frac{k}{a^{2}} &=&-\frac{1}{%
			M_{P}^{2}}\Bigg[p_{\upphi }+p_{c}^{(\upphi )}+p_{rad}+p_{c}^{(rad)}
		+p_{DM}+p_{c}^{(DM)}\Bigg],
	\end{eqnarray}%
	respectively. In the following we take $M_{P}=1$, according to $\hbar=G=1$. The equation for the radiation particle number can be
	immediately converted into an equation giving the variation of the
	temperature of the Universe as
	\begin{equation}
		\dot{T}+HT=\frac{\Upgamma }{96\uppi \upzeta \left( 3\right) }\frac{1}{T^{2}}%
		\frac{\uprho _{\upphi }}{m_{\upphi }},
	\end{equation}%
	while the radiation creation pressure becomes
	\begin{equation}
		p_{c}^{(rad)}=-\frac{\Upgamma \uppi ^{4}}{135\upzeta (3)}\frac{T}{H}\frac{\uprho
			_{\upphi }}{m_{\upphi }}.
	\end{equation}
	
	Eq. (\ref{KG}) for the energy density variation of the scalar field takes
	the form of a generalized Klein-Gordon equation, given by
	
	\begin{equation}
		\ddot{\upphi}+3H\dot{\upphi}+V^{\prime }\left( \upphi \right) =-\Upgamma \dot{\upphi}%
		\frac{\uprho _{\upphi }}{m_{\upphi }n_{\upphi }}.
	\end{equation}
	
	\section{Warm inflation in the presence of a coherent scalar field}\label{sect4}
	
	In the following we will investigate a simple warm inflationary cosmological model, obtained
	under the assumption that the inflationary scalar field is a homogeneous
	field, oscillating with a frequency $m_{\upphi }$. Such a field can
	be interpreted as a coherent wave of scalar \textquotedblleft
	particles\textquotedblright\, having zero momenta, and with the corresponding
	particle number density given by
	\begin{equation}
		n_{\upphi }=\frac{\uprho _{\upphi }}{m_{\upphi }}.
	\end{equation}%
	
	Therefore, this model corresponds to $n_{\upphi }$ oscillators, having the same frequency $%
	m_{\upphi }$, with all scalar particles oscillating coherently with the same phase. Hence, the warm inflationary evolution can be
	described as determined by a single homogeneous scalar wave $\upphi (t)$. For this type of
	scalar field it follows that $p_{\upphi }=0$, giving
	\begin{equation}
		V\left(
		\upphi \right)=\frac{1}{2}\dot{\upphi}^{2},
	\end{equation}
	and
	\begin{equation}
		\uprho _{\upphi }=\dot{\upphi}^{2},
	\end{equation}
	respectively.
	
	Under this choice of the scalar field, the set of cosmological equations describing the warm
	inflationary dynamics takes the form
	\begin{equation}
		\dot{\uprho}_{\upphi }+3H\uprho _{\upphi }=-\Upgamma \uprho _{\upphi },  \label{s1}
	\end{equation}
	\begin{equation}
		\dot{T}+HT=\frac{\Upgamma }{96\uppi \upzeta \left( 3\right) m_{\upphi }}\frac{1}{%
			T^{2}}\uprho _{\upphi },  \label{s2}
	\end{equation}
	\begin{equation}
		\dot{\uprho}_{DM}+3H\uprho _{DM}=\frac{\Upgamma }{2}\frac{m_{DM}}{m_{\upphi }}\uprho_{\upphi}, \label{s3}
	\end{equation}
	
	\begin{equation}
		3\frac{\dot{a}^{2}}{a^{2}}+3\frac{k}{a^{2}}=\frac{1}{M_{P}^{2}}\left( \uprho
		_{\upphi }+\frac{8\uppi ^{5}}{15}T^{4}+\uprho _{DM}\right) .  \label{s4}
	\end{equation}
	
	The system of differential equations (\ref{s1})-(\ref{s4}) represents a system of four ordinary differential equations with four unknowns $\{
	a,\uprho _{\upphi },\uprho _{DM},T\} $. It must be integrated with the
	initial conditions $a(0)=a_{0}$, $\uprho _{\upphi }(0)=\uprho _{\upphi0}$, $\uprho_{DM}(0)=\uprho_{DM0}$, $T(0)=T_{0}$.
	
	In order to simplify the mathematical formalism we introduce a set ot of dimensionless variables $\{ \uptau ,A,r_{\upphi },r_{DM},\uptheta \} $, defined as
	\begin{eqnarray}\label{mphi}
		\uptau &=&\Upgamma t, \; A=\Upgamma a, \; r_{\upphi }=\frac{1}{3M_{P}^{2}\Upgamma ^{2}}%
		\uprho _{\upphi },\nonumber\\
		r_{DM}&=&\frac{1}{3M_{P}^{2}\Upgamma ^{2}}\uprho _{DM}, \; \uptheta=\left(
		\frac{M_{P}^{2}\Upgamma ^{2}}{32\uppi \upzeta (3)m_{\upphi }}\right)
		^{-1/3} T .
	\end{eqnarray}
	
	Then the system of equations (\ref{s1})-(\ref{s4}) takes the following
	dimensionless form,
	\begin{equation}\label{scfield}
		\frac{dr_{\upphi }}{d\uptau }+3\frac{1}{A}\frac{dA}{d\uptau }r_{\upphi }=-r_{\upphi },
	\end{equation}
	\begin{equation}\label{rad}
		\frac{d\uptheta }{d\uptau }+\frac{1}{A}\frac{dA}{d\uptau }\uptheta =\frac{r_{\upphi }}{%
			\theta ^{2}},
	\end{equation}
	\begin{equation}\label{DM}
		\frac{dr_{DM}}{d\uptau }+3\frac{1}{A}\frac{dA}{d\uptau }r_{DM}=\frac{\uplambda}{2} r_{\upphi },
	\end{equation}
	\begin{equation}
		\frac{1}{A^{2}}\left( \frac{dA}{d\uptau }\right) ^{2}+\frac{k}{A^{2}}=r_{\upphi
		}+\upalpha \uptheta ^{4}+r_{DM},
	\end{equation}%
	where we have denoted
	\begin{equation}
		\uplambda =\frac{m_{DM}}{m_{\upphi }},\upalpha =\frac{8\uppi ^{5}}{45M_{P}^{2}\Upgamma
			^{2}}\left( \frac{M_{P}^{2}\Upgamma ^{2}}{32\uppi \upzeta (3)m_{\upphi }}\right)
		^{4/3}.
	\end{equation}
	
	We have chosen the numerical values for the initial conditions used for the numerical integration of the cosmological evolution equations: $A_{0}=0.1$, $\uptheta_{0}=0.0001$, $r_{DM0}=0.005$, $r_{\upphi0}=20$. For the dimensionless parameters of the model we adopt the values $\uplambda=6$ and $\upalpha=10$, whereas for the geometry of the primordial Universe we successively consider all three possibilities - flat, open and closed, respectively.
	
\begin{figure}[H]
		\centering
		\includegraphics[scale=0.70]{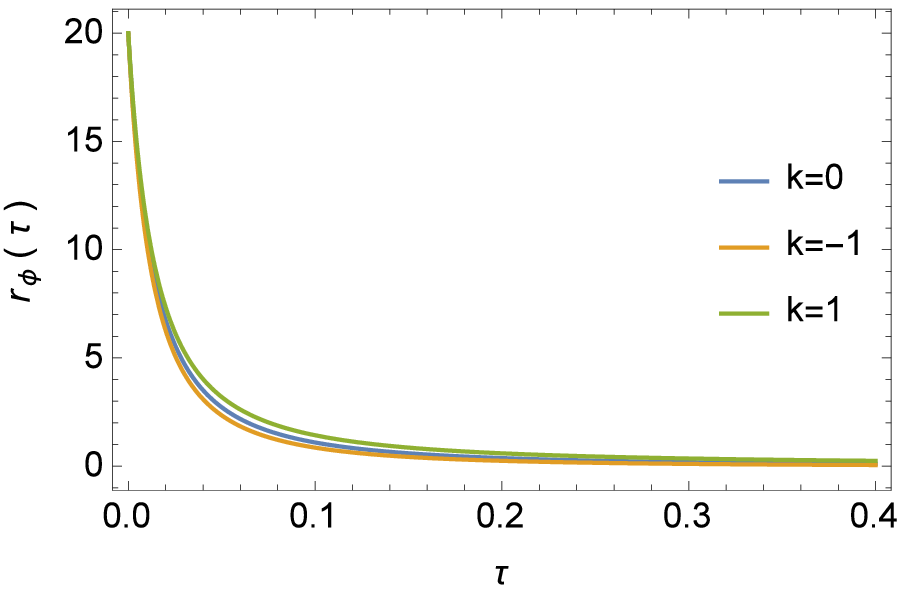}
		\includegraphics[scale=0.70]{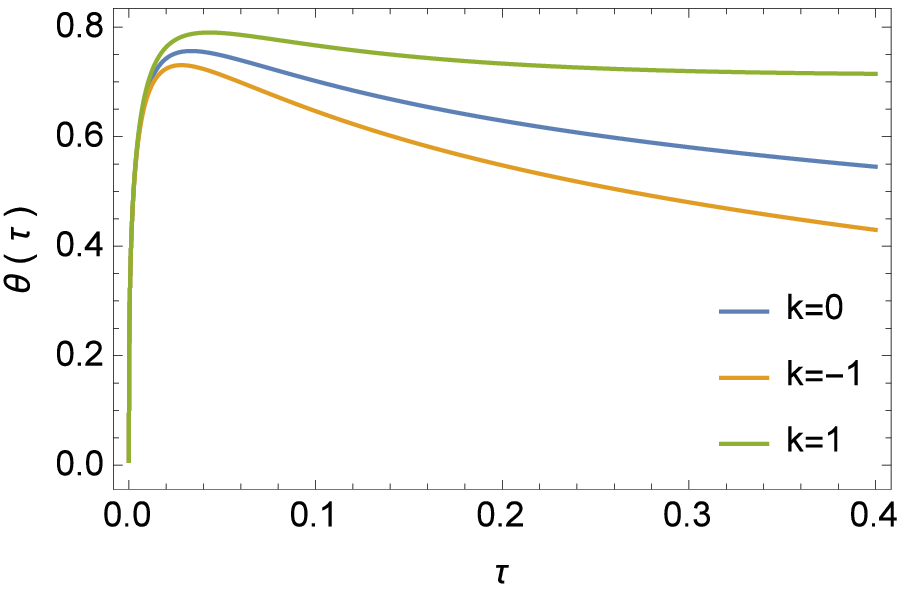}
		\caption{Time variation in a warm inflationary regime of the energy density of the scalar field  $r_{\upphi}(\uptau)$ (left panel) and of the temperature $\uptheta (\uptau)$ (right panel)  in a Universe having an initial flat geometry, $k=0$ (blue curve), an open geometry, $k=-1$ (orange curve), and a closed geometry, $k=1$ (green curve), respectively. }
		\label{fig1}
	\end{figure}

	The variation of the energy density of the scalar field, and of the temperature of the radiation fluid are represented in Fig.~\ref{fig1}. As one can see from the left panel of the Figure, the scalar field decays rapidly due to matter creation, but there is no dependence on the initial geometry of the Universe, therefore all cases have a very similar decay. In the temperature evolution of the radiation fluid (right panel of Fig~\ref{fig1}) a dependence can be seen between the maximum reheating temperature and the initial geometry of the Universe. The reheating temperature, which is the maximum temperature of the Universe acquired by the end of inflation, is slightly reduced by a shift from an initially flat Universe to an open one and slightly increased considering a closed initial Universe, which moreover seems to contain a remnant component in a later Universe.
	
\begin{figure}[H]
		\centering
		\includegraphics[scale=0.70]{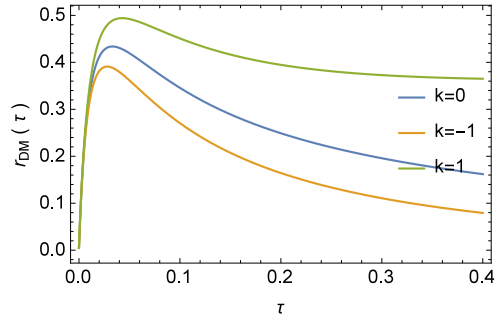}
		\caption{Time variation in a warm inflationary regime of the dark matter energy density $r_{DM}(\uptau)$ in a Universe having an initial flat geometry, $k=0$ (blue curve), an open geometry, $k=-1$ (orange curve), and a closed geometry, $k=1$ (green curve), respectively. }
		\label{fig3}
	\end{figure}

	The same dependence on the initial curvature of the Universe can be observed in the time variation of the energy density of the newly created dark matter, presented in Fig.~\ref{fig3}, with the maximum values of the dark matter density occurring at roughly similar times as compared to the maximum values of the temperature of the radiation fluid. Analogous to the temperature function, the closed geometry also shows some remnants of dark matter particles produced during the warm inflationary era.     
	
\begin{figure}[H]
		\centering
		\includegraphics[scale=0.70]{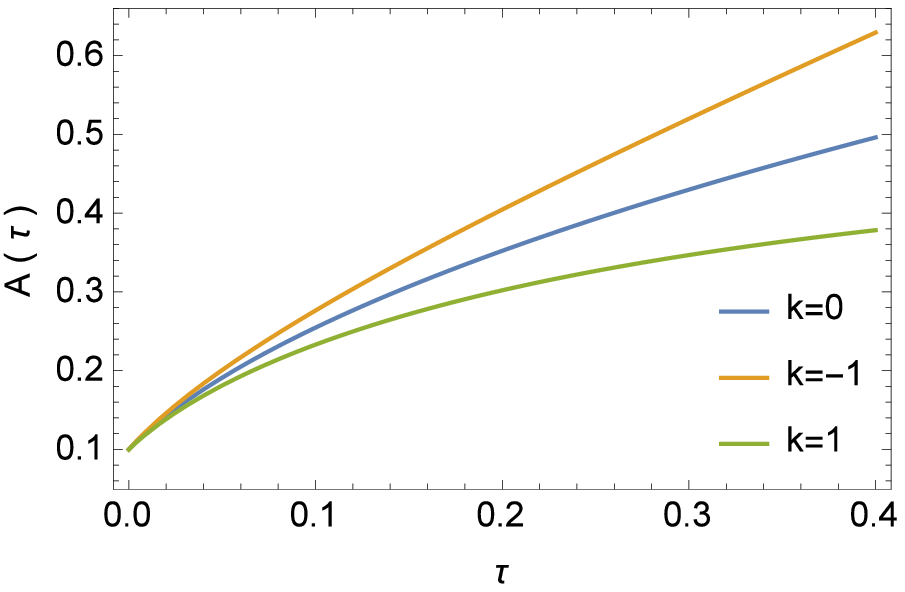}
		\includegraphics[scale=0.70]{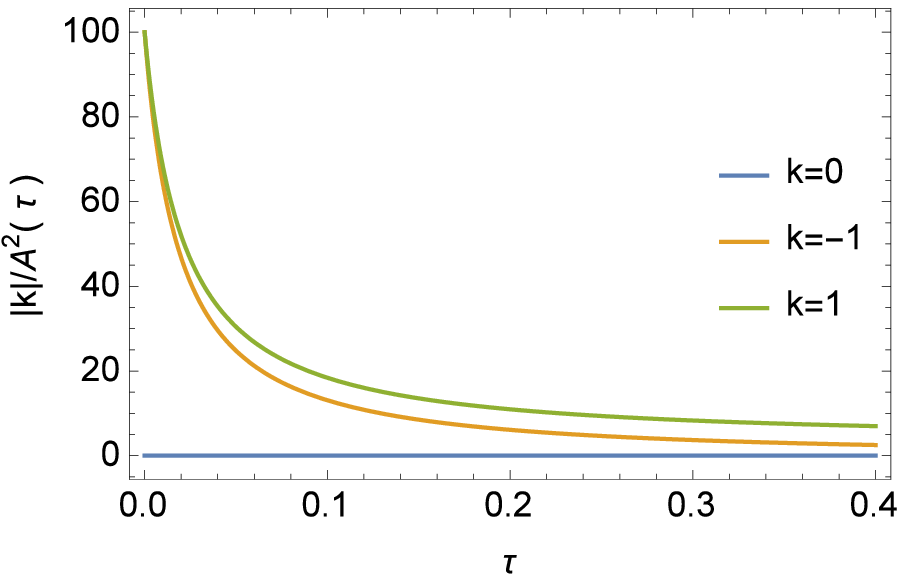}
		\caption{Time variation in a warm inflationary regime of the scale factor $A(\uptau)$ (left panel), and of the curvature term $|k|/A^2(\uptau)$ (right panel) in a Universe having a flat geometry, $k=0$ (blue curve), an open geometry, $k=-1$ (orange curve), and a closed geometry, $k=1$ (green curve), respectively. }
		\label{fig4}
	\end{figure}

	The time variations of the scale factor and of the curvature terms $|k|/A^2(\uptau)$ are depicted in Fig.~\ref{fig4}. The expansion rate also depends on the initial curvature, with the slowest expansion corresponding to the closed cosmological geometry. While the curvature terms reach quickly the zero value in the open and flat cases, in the present model this is not the case for an initial closed Universe, whose curvature may reach the zero value only in the asymptotic limit of very large times. 
	
	\section{Discussions and final remarks}\label{sect5}
	
	The present paper aims to address from an open thermodynamical system perspective the problem of the creation of particles in a warm inflationary scenario by considering a three-component dynamical system composed of a scalar field, radiation, and dark matter, respectively. Moreover, we have assumed that the Universe may have had an initial curvature at the moment of its very beginning, and we have explored the role this curvature may have had on the evolution of the physical properties of the cosmological system.  The work conducted in this paper intends to enlarge the approach of \citet{Harko2020}, by including a dark matter and a curvature component into the cosmological field equations of the  warm inflationary formalism.
	
	In our study we have considered the early Universe as an open thermodynamic systems in which entropy and particle creation occurs \citep{Prigogine1988}. The time evolution of the dynamically interacting cosmological fluid consisting of a scalar field, radiation and dark matter has been investigated in the curved FLRW geometry, with the effects of the geometric curvature terms fully taken into account.
	
	Some important features of this model can be observed from the behaviour of the physical and geometrical parameters, as obtained in the previous Section. As the scalar field decays into the newly created radiation and dark matter particles (see Fig.~\ref{fig1}), the temperature of the early Universe is bound to increase, in the case of a flat, $k=0$, open, $k=-1$, or closed, $k=1$, geometry, from a zero value to a maximum value. After reaching its maximum, the temperature decreases, due to the accelerated expansion of the Universe.
	
	The maximum value of the photon gas temperature, which can be considered as the reheating temperature, is an important cosmological parameter. In order to obtain the reheating temperature values, we mention first that the evolution equation of the scalar field energy density, Eq.~ (\ref{scfield}), can be integrated exactly to give the expression
	\begin{equation}
		r_{\upphi}=r_{\upphi0} \, \frac{e^{-\uptau}}{A^3},
	\end{equation}
	where $r_{\upphi0}$ is an arbitrary constant of integration.  From Eq.~(\ref{rad}) we obtain the maximum value of the temperature as
	\begin{equation}
		\uptheta_{max}=\left.\left[\frac{r_{\upphi}}{\left(1/A\right)\left(dA/d\uptau\right)}\right]^{1/3}\right|_{\uptau=\uptau_{max}}.
	\end{equation}
	
	From Fig.~\ref{fig1} one can obtain the reheating maximum temperatures $\uptheta_{max}$,  as 
	$\uptheta_{max}^{(-1)}(\uptau_{max}=0.02693)=0.7258$ for the open geometry, $\uptheta_{max}^{(0)}(\uptau_{max}=0.03293)=0.7585$ for the flat geometry, and $\uptheta_{max}^{(+1)}(\uptau_{max}=0.04133)=0.7912$ for the closed geometry.
	
	The energy density of the dark matter particles $r_{DM}(\uptau)$ has a similar behavior to that of the temperature. The maximum value of the dark matter energy density can be obtained from Eq.~(\ref{DM}) as,
	\begin{equation}
		r_{DM}^{(max)}=\left.\frac{\uplambda r_{\upphi}}{6\left(1/A\right)\left(dA/d\uptau\right)}\right|_{\uptau=\uptau_{max}}.
	\end{equation}
	
	For an open Universe,
	$\left.r_{DM}^{(max)}\right|_{k=-1}\left(\uptau_{max}=0.02767\right)=0.3872$,
	for a flat Universe $\left.r_{DM}^{(max)}\right|_{k=0}\left(\uptau_{max}=0.03257\right)=0.4316$ and for a closed Universe,\\ $\left.r_{DM}^{(max)}\right|_{k=1}\left(\uptau_{max}=0.04237\right)=0.4916$.
	It is interesting to note that a decrease in $\uplambda$ reduces the peak value of the dark matter density, so that for a value of $\uplambda_{critic}=0.1$, the maximum disappears. Taking into consideration the definition of $\uplambda$,  a relationship between the mass of the dark matter particles and the mass of the scalar field particle can be found as $\uplambda_{critic}=0.1=m_{DM}/m_{\upphi}$, and therefore $m_{DM}\geq 0.1 \, m_{\upphi}$.
	
	We have also investigated, in our analysis, the behavior of the curvature terms during the warm inflationary period. For an open Universe, the curvature term $|k|/A^2(\uptau)$ heads towards the zero value, indicating that the Universe reaches the flat geometry in a very short time interval. Nevertheless, in a closed geometry, the curvature term does not approach zero, which implies the necessary existence of a remnant curvature in the late Universe. Moreover, the closed geometry scenario imposes a strong requirement for the initial conditions of the Universe, so that real solutions of the field equations can be found, namely, 
	\begin{equation}
		\frac{1}{3} \, \left.(r_{\upphi} + \upalpha \uptheta^4 + r_{DM} )\right|_{\uptau=\uptau_0}  > 1. 
	\end{equation}
	For zero initial values of the temperature of the radiation fluid, and for the density of the dark matter, it follows that $\left.r_{\upphi}\right|_{\uptau=\uptau_0}$ must be strictly greater than 3. Hence, an initially closed Universe can not exist, unless there is a fine-tuning with regard of the initial conditions.
	
	According to the standard Big Bang model, the reheating regime must end at the reheating time $t_{reh}=10^{-18}$ s, which signifies the time at which the reheating temperature is attained, with the corresponding reheating temperature being of the order of $T_{reh}<10^{15}$ GeV. Therefore, $\Upgamma$ can be computed from the maximum of $\uptau$, its value depending directly upon the geometry of the early Universe, $ \Upgamma = \uptau_{max}^{(k=\{0; \pm 1\})}/ t_{reh}.$
	
	Moreover, we can aquire some information about the mass of the dark matter particles considering Eq. (\ref{mphi}) where we substitute $\Upgamma$ with the previously presented form, from where
	\begin{eqnarray} \label{mdm}
		m_{DM}&=& \lambda \, \frac{M_{P}^2}{32 \uppi \upzeta(3)}%
		\frac{\uptau_{max}^2}{t_{reh}^2} \frac{\uptheta_{max}^3}{T_{reh}^3} .
	\end{eqnarray}
	Therefore, the dark matter particle mass has the following numerical value at the end of the reheating regime:
	\begin{eqnarray}
		m_{DM}&=& 5.34\times 10^{-22} \times \uplambda \times \uptau_{max}^2 \times 	\uptheta_{max}^3 \\ \nonumber 
		&\times& \left(\frac{t_{reh}}{10^{-18} \; \rm{s}}\right)^{-2} \times \left(\frac{T_{reh}}{10^{15} \; \rm{GeV}} \right)^{-3} \rm{GeV}.
	\end{eqnarray}
	
	If we choose for the mass coefficient the value $\uplambda=6$ and for the maximum of the photon gas temperature the value $\uptheta_{max}=0.759$, its peak occurring at the time $\uptau_{max}=0.034$, we obtain for the dark matter mass the value $m_{DM}=1.62\times 10^{-15}\; \rm{eV}$. Moreover, by the same numerical selection of $\uplambda,\; \uptheta_{max}$ and $ \uptau_{max}$, yet increasing the reheating temperature by two orders of magnitude, the mass reaches $10^{-21}\; \rm{eV}$. This result is comparable to the mass of an ultralight axion-like particle  measured by gravitational weak lensing in the study of \citet{Dentler}, in which the dark matter particle is determined to have a mass of the order of $10^{-22}\; \rm{eV}$.
	
	From the analysis of the static Bose-Einstein Condensate dark matter halos it follows that the condensate dark matter satisfies a mass-galactic radius relation of the form \citep{HaBo}
	\begin{eqnarray} \label{mass}
		m &=&\left( \frac{\pi ^{2}\hbar ^{2}a}{GR^{2}}\right) ^{1/3}\approx
		6.73\times 10^{-2}\times \left[ a\left( \mathrm{fm}\right) \right]
		^{1/3}%
		\left[ R\;\mathrm{(kpc)}\right] ^{-2/3}\;\mathrm{eV},
	\end{eqnarray}
	where $a$ is the scattering length of the particle, and $R$ is the galactic radius. By introducing the ratio $\sigma _m=\sigma /m$ of the self-interaction cross section $\sigma =4\pi a^2$, and of the dark matter particle mass $m$, then the mass of the dark matter particle can be obtained from Eq.~(\ref{mass}) as \citet{Blaga2015}
	\begin{equation}
		m=\left(\frac{\pi ^{3/2}\hbar ^2}{2G}\frac{\sqrt{\sigma _m}}{R^2} \right)^{2/5}.
	\end{equation}
	
	By assuming for $\sigma _m$ a value of the order of $\sigma _m=1.25\; %
	\mathrm{cm^2/g}$ \citep{Blaga2015}, we obtain for the mass of the dark matter particle the constraint upper limit of the order
	\begin{eqnarray}\label{massg}
		m&<&3.1933\times10^{-37}\left(\frac{R}{10\;\mathrm{kpc}}\right)^{-4/5}\times
		\left(\frac{\sigma _m}{1.25\;\mathrm{cm^2/g}}\right)^{1/5}\;\mathrm{g} \\ \nonumber
		&=& 0.1791\times\left(\frac{R}{10\;\mathrm{kpc}}\right)^{-4/5} \times 
		\left(\frac{\sigma_m}{1.25\;\mathrm{cm^2/g}}\right)^{1/5}\;\mathrm{meV}. 
	\end{eqnarray}
	
	By assuming that the self-interaction cross section of the dark matter particle takes values in the range $\sigma_m\in(0.00335\;\mathrm{cm^2/g},0.0559\;\mathrm{cm^2/g})$, we obtain for the possible range of the mass of the dark matter as given by
	\begin{eqnarray} \label{massg1}
		m&\approx&\left(9.516\times 10^{-38}-1.670\times
		10^{-37}\right)\left(\frac{R}{10\;\mathrm{kpc}}\right)^{-4/5}\;\mathrm{g} \\ \nonumber
		&=& \left(0.053-0.093\right)\left(\frac{R}{10\;\mathrm{kpc}}\right)^{-4/5}\;\mathrm{meV}. 
	\end{eqnarray}
	
\begin{figure}[H]
		\centering
		\includegraphics[scale=0.5]{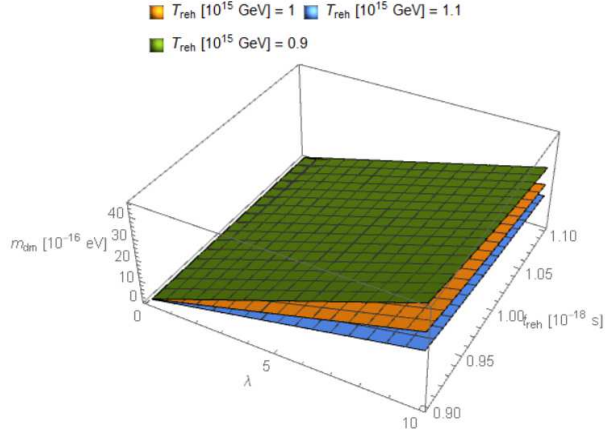}
		\caption{Dark matter mass variation as a function of mass coefficient and reheating time when  $t_{reh}=10^{-18}\; \rm{s}$ and $T_{reh}=10^{15}\; \rm{GeV}$ (orange plane), when there is a 10 percent increase of the reheating temperature value (blue plane), and at a 10 percent decrease of the reheating temperature value, respectively (green plane). }
		\label{3d}
	\end{figure}

	In order to find the dark matter mass $m_{DM}$ in the upper limit outlined in \citet{Blaga2015} in the form of Eq.~(\ref{massg1}), the parameters $\uplambda, \; t_{reh}, \; T_{reh}$ from Eq.~(\ref{mdm}) must be adjusted accordingly. A three-dimensional representation of the dark matter mass as a function of the mass coefficient and reheating time is illustrated in Fig.~\ref{3d}. One can clearly see the proportionality of $m_{DM}$ to the mass coefficient, as the dark matter mass grows with the increase of $\uplambda$. Moreover, there is an inverse proportionality of $m_{DM}$ to the reheating time and temperature, so that although the system accumulates mass, which is of the order of $10^{-15}\; \rm{eV}$, the reheating time and temperature diminish in value.
	
\begin{figure}[H]
		\centering
		\includegraphics[scale=0.5]{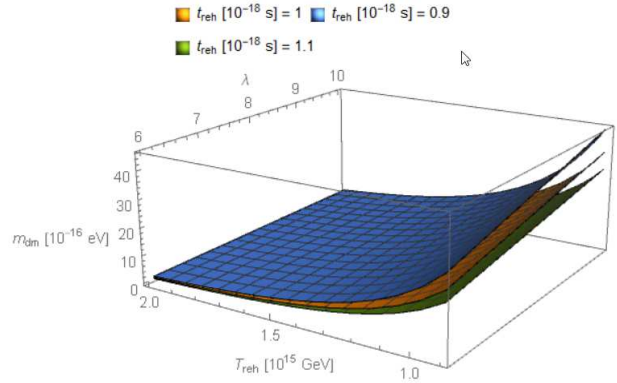}
		\caption{Dark matter mass variation as a function of mass coefficient and reheating temperature when $t_{reh}=10^{-18}\; \rm{s}$ and $T_{reh}=10^{15}\; \rm{GeV}$ (orange plane), when there is a 10 percent decrease of the reheating time value (blue plane), and at a 10 percent increase of the reheating time value, respectively (green plane). }
		\label{3d1}
	\end{figure}

	The representation of the dark matter mass as a function of $\uplambda$ and of the photon gas temperature can be seen in Fig.~\ref{3d1}. It is notable that by increasing the reheating temperature from $0.9 \times 10^{15}\; \rm{GeV}$ to $2.0 \times 10^{15}\; \rm{GeV}$, the mass function reveals a decreasing behaviour at a steep rate for high enough mass coefficients $\uplambda$. The dark matter mass enlarges as the reheating time is decreased along with the temperature by ten percent of their original value.
	
	The present paper has only considered a warm inflationary scenario in the presence of a homogeneous coherently oscillating scalar field, which has given a simple cosmological model with an interacting three-components cosmological fluid, whose evolution is described by using the thermodynamics of open systems in the presence of matter creation. Additional forms of the scalar field will be taken into consideration in a future study.

	\makeatletter
	\def\@biblabel#1{}
	\makeatother
	
	\received{\it *}

\begin{thebibliography}{99}
		\bibitem[Albrecht \etal, 1982]{Albrechts1982}  Albrecht A., Steinhardt P.J. , Turner M.S., Wilczek F.: 1982, \textit{Phys. Rev. Lett.} \textbf{48}, 1437.
		\bibitem[Albrecht and Steinhardt, 1982]{Albrecht1982}  Albrecht A., Steinhardt P.J.: 1982, \textit{Phys. Rev. Lett.} \textbf{48}, 1220.
		\bibitem[Antusch \etal, 2015]{Antusch2015}  Antusch S., Nolde D., Orani S.: 2015, \textit{J. Cosmol. Astropart. Phys.} \textbf{06}, 009.
		\bibitem[Bastero-Gil \etal, 2011]{Bastero2011}  Bastero-Gil M., Berera A., Ramos R.O.: 2011, \textit{J. Cosmol. Astropart. Phys.} \textbf{1107}, 030.
		\bibitem[Berera and Fang, 1995]{BereraFang}  Berera A., Fang L.-Z.: 1995, \textit{Phys. Rev. Lett.} \textbf{74} 1912.
		\bibitem[Berera \etal, 2009]{BereraMoss}  Berera A., Moss I. G. and Ramos R. O.: 2009, \textit{Reports on Progress in Physics} \textbf{72}, 026901.
		\bibitem[Berera, 1996]{Berera1996}  Berera A.: 1996, \textit{Phys. Rev.} \textbf{D 54}, 2519.
		\bibitem[Berera, 1995]{Berera1995}  Berera A.: 1995, \textit{Phys. Rev. Lett.} \textbf{75}, 3218.
		\bibitem[Böhmer and T. Harko, 2007]{HaBo} Böhmer C. G., Harko T.: 2007, \textit{JCAP} {\bf 0706}, 025.
		\bibitem[Calvão \etal, 1992]{Calvao1992}  Calvão M., Lima J., Waga I.: 1992, \textit{Phys. Lett.} \textbf{B 162}, 223.
		\bibitem[Chakraborty and Saha, 2014]{ChakrabortySaha}  Chakraborty S., Saha S.: 2014, \textit{Phys. Rev.} \textbf{D 90}, 123505.
		\bibitem[Chakraborty, 2014]{Chakraborty2014}  Chakraborty S.: 2014, \textit{Phys. Lett.}  \textbf{ B 732}, 81.
		\bibitem[De Oliveira and Jorás, 2001]{Oliveira2001}  De Oliveira H.P., Jorás S.E.: 2001, \textit{Phys.Rev.} \textbf{D 64}, 063513.
		\bibitem[Dentler \etal, 2022]{Dentler} Dentler M., Marsh D.J.E., Hložek R., Laguë A., Rogers K.K., Grin D.:2022, \textit{Monthly Notices of the Royal Astronomical Society} \textbf{515}, arXiv:2111.01199.
		\bibitem[Gleisler and Ramos, 1994]{Gleisler1994}  Gleiser M. and Ramos R. O.: 1994, \textit{Phys. Rev.} \textbf{D 50}, 2441.
		\bibitem[Guth, 1981]{Guth1981}  Guth A.: 1981, \textit{Phys. Rev.} \textbf{D 23}, 347.
		\bibitem[Hall \etal, 2004]{Hall2004}  Hall L.M.H., Moss I.G., Berera A.: 2004, \textit{Phys. Rev.} \textbf{D 69}, 083525.
		\bibitem[Harko and Sheikhahmadi, 2020]{Harko2020} Harko T. and Sheikhahmadi H.: 2020, \textit{Physics of the Dark Universe} \textbf{28}, 100521.
		\bibitem[Harko and Sheikhahmadi, 2021]{Harko2021}  Harko T. and Sheikhahmadi H.: 2021, \textit{Eur. Phys. J. C}, \textbf{81}, 165.
		\bibitem[Harko \etal, 2015]{Blaga2015} Harko T., Liang P., Liang S.-D., and Mocanu G.:2015, \textit{JCAP} {\bf 11},  027.
		\bibitem[Harko and Lobo, 2013]{Harko2013}  Harko T., Lobo F.S.N.: 2013, \textit{Phys. Rev.} \textbf{D 87}, 044018.
		\bibitem[Harko, 2014]{Harko2014}  Harko T.: 2014, \textit{Phys. Rev.} \textbf{D 90}, 044067.
		\bibitem[Kofman \etal, 1994]{Kofman1994}  Kofman L., Linde A., Starobinsky A.A.: 1994, \textit{Phys. Rev. Lett.} \textbf{73}, 3195.
		\bibitem[Liddle and Barrow, 1994]{Liddle1994}  Liddle P. P. A. R. and Barrow J. D.: 1994, \textit{Phys. Rev.} \textbf{D 50}, 7222.
		\bibitem[Lima and Baranov, 2014]{Lima2014}  Lima J.A.S., Baranov I.: 2014, \textit{Phys. Rev.} \textbf{D 90}, 043515.
		\bibitem[Lima \etal, 2016]{Lima2016}  Lima J.A.S., Basilakos S., Solà J.: 2016, \textit{Eur. Phys. J.} \textbf{C 76}, 228.
		\bibitem[Linde, 1982]{Linde1982}  Linde A.: 1982, \textit{Phys. Lett.} \textbf{B 108}, 389.
		\bibitem[Linde, 1983]{Linde1983}  Linde A.: 1983, \textit{Phys. Lett.} \textbf{B 129}, 177.
		\bibitem[Linde, 1994]{Linde1994}  Linde A.: 1994, \textit{Phys. Rev.} \textbf{D 49}, 748.
		\bibitem[Modak and Singleton, 2012]{Modak2012}  Modak S.K., Singleton D.: 2012, \textit{Phys. Rev.} \textbf{D 86}, 123515.
		\bibitem[Mukhanov, 2005]{Mukhanov2005} Mukhanov V.: 2005, \textit{Physical Foundations of Cosmology, Cambridge University Press, Cambridge, UK}.
		\bibitem[Nojiri \etal, 2017]{Nojiri2017} Nojiri S., Odintsov S.D., Oikonomou V.K.: 2017, \textit{Phys. Rep.} \textbf{692}, 1.
		\bibitem[Nunes and Pan, 2016]{Nunes2016}  Nunes R.C., Pan S.: 2016, \textit{Mon. Not. R. Astron. Soc.} \textbf{459}, 673.
		\bibitem[Pigozzo \etal, 2016]{Pigozzo2016} Pigozzo C., Carneiro S., Alcaniz  J.S. , Borges H.A., Fabris J.C.: 2016, \textit{J. Cosmol. Astropart. Phys.} \textbf{05}, 022.
		\bibitem[Pinto \etal, 2022]{Harko2022} Pinto M. A. S., Harko T., and Lobo F. S. N.:2022, \textit{Phys. Rev.} \textbf{D 106}, 044043.
		\bibitem[Prigogine \etal, 1988]{Prigogine1988}  Prigogine I., Geheniau J., Gunzig E., Nardone P.: 1988, \textit{Proc. Natl. Acad. Sci.} \textbf{85}, 7428.
		\bibitem[Qiang \etal, 2007]{Qiang2007}  Qiang Y., Zhang T.-J., Yi Z.-L.: 2007, \textit{Astrophys. Space Sci.} \textbf{311}, 407. 
		\bibitem[Stewart, 2002]{Stewart2002} Stewart E. D.: 2002, \textit{Phys. Rev.} \textbf{D 65}, 103508.
		\bibitem[Su \etal, 2017]{Su2017}   Su J., Harko T., Liang S.-D.: 2017, \textit{Adv. High Energy Phys.} \textbf{2017}, 76502398.
		\bibitem[Zimdahl \etal, 1996]{Zimdahl1996}  Zimdahl W., Triginer J., Pavon D.: 1996, \textit{Phys. Rev.} \textbf{D 54}, 6101.
		
	\end{thebibliography}
\end{document}